\documentclass[preprint,10pt]{aastex}
\usepackage[utf8]{inputenc}

\newcommand{\numManual}{37}     
\newcommand{\numBlendSets}{6}   

\newcommand{\numEBs}{31}        
\newcommand{\numEBstext}{Thirty-one}
\newcommand{\numEBsnew}{20}     
\newcommand{\numTargs}{1951}    
\newcommand{\numEngAps}{128}    
\newcommand{\fracEBs}{1.6\%}    

\newcommand{\kepler}{{\sl Kepler~}}

\begin{document}

\title{Kepler Eclipsing Binary Stars. V. Identification of \numEBs\ Eclipsing Binaries in the K2 Engineering Data-set}
\author{Kyle E. Conroy}
\affil{Department of Physics and Astronomy, Vanderbilt University, VU Station B 1807, Nashville, TN 37235}

\author{Andrej Pr\v{s}a}
\affil{Department of Astrophysics and Planetary Sciences, Villanova University, 800 E Lancaster Ave, Villanova, PA 19085}

\author{Keivan G. Stassun}
\affil{Department of Physics and Astronomy, Vanderbilt University, VU Station B 1807, Nashville, TN 37235}
\affil{Department of Physics, Fisk University, Nashville, TN 37208}

\author{Steven Bloemen}
\affil{Dept. of Astrophysics, IMAPP, Radboud University Nijmegen, PO Box 9010, 6500 GL, Nijmegen, The Netherlands; Instituut voor Sterrenkunde}

\author{Mahmoud Parvizi}
\affil{Department of Physics and Astronomy, Vanderbilt University, VU Station B 1807, Nashville, TN 37235}

\author{Billy Quarles}
\affil{NASA Ames Research Center, M/S 244-30, Moffett Field, CA 94035, USA}

\author{Tabetha Boyajian}
\affil{Department of Astronomy, Yale University, New Haven, CT 06511, USA}

\author{Thomas Barclay}
\affil{NASA Ames Research Center, M/S 244-30, Moffett Field, CA 94035, USA}
\affil{Bay Area Environmental Research Institute, 596 1st Street West, Sonoma, CA 95476, USA}

\author{Avi Shporer}
\affil{Division of Geological and Planetary Sciences, California Institute of Technology, Pasadena, CA 91125, USA}
\affil{Jet Propulsion Laboratory, California Institute of Technology, 4800 Oak Grove Drive, Pasadena, CA 91109, USA ; NASA Sagan Fellow}

\author{David W.~Latham}
\affil{Harvard-Smithsonian Center for Astrophysics, 60 Garden St
reet, Cambridge, MA 02138}

\author{Michael Abdul-Masih}
\affil{Department of Astrophysics and Planetary Sciences, Villanova University, 800 E Lancaster Ave, Villanova, PA 19085}


\begin{abstract}

Over 2500 eclipsing binaries were identified and characterized from the ultra-precise photometric data provided by the \kepler space telescope. \kepler is now beginning its second mission, K2, which is proving to again provide ultra-precise photometry for a large sample of eclipsing binary stars. In the \numTargs\ light curves covering 12 days in the K2 engineering data-set, we have identified and determined the ephemerides for \numEBs\ eclipsing binaries that demonstrate the capabilities for eclipsing binary science in the upcoming campaigns in K2. Of those, \numEBsnew\ are new discoveries. We describe both manual and automated approaches to harvesting the complete set of eclipsing binaries in the K2 data, provide identifications and details for the full set of eclipsing binaries present in the engineering data-set, and discuss the prospects for application of eclipsing binary searches in the K2 mission.
\end{abstract}

\section{Introduction\label{sec:intro}}

The \kepler satellite \citep{Batalha10} observed over 150,000 stars in its original mission which acquired over 4 years of high-precision photometry.  This data-set was followed by a large effort to study the eclipsing binary (EB) population in the \emph{Kepler} field, resulting in the detection and characterization of over 2500 EB stars \citep{Prsa11,Slawson11,KirkPrep}, the measurements of eclipse timing variations \citep{Conroy14,OroszPrep}, and the discovery of several circumbinary planets \citep{Kepler16,Kepler34}.

Now that \kepler has transitioned to its re-purposed mission, K2, it is providing 80 days of continuous high-precision photometry across each of 10 fields in the ecliptic plane, once again giving great scientific opportunity to identify and characterize EBs \citep{prsa2014}. Although the photometric precision compared to the original Kepler mission is expected to be slightly lower due to a decrease in pointing accuracy, K2 is still expected to obtain data an order of magnitude better than is possible from the ground.  With the upcoming TESS mission, EBs identified in K2 will become prime targets for further follow-up -- allowing us to extend the time baseline and continue searching for triple systems (stellar and substellar) through eclipse timing variations and searching for transiting events.

Nonetheless, it is important to assess both the potential and the challenges of harvesting EBs from the new K2 data. In this paper, we utilize the first publicly available data-set from K2---the engineering data-set--- to perform a cursory look at the EB identification methods as applied to K2. In Sec.~\ref{sec:data} we describe the K2 data that we use and the data-level processing of the K2 light curves. Sec.~\ref{sec:results} we present the manual and automated methods that we employ to identify and classify the EBs in the K2 data-set along with their ephemerides. We conclude in Sec.~\ref{sec:summary} with a brief summary and a brief discussion of prospects for EB science in the full upcoming K2 mission.

\section{K2 Data and Processing\label{sec:data}}

Unlike the main \kepler mission that focused on a predetermined set of targets within the fixed field of view, the targets for each K2 campaign are solicited from the community, with $\sim$10,000 long-cadence and $\sim$100 short-cadence targets selected for observations from each field \citep{howell2014}. The \kepler Eclipsing Binary Working Group contributes a selection of science targets based on a cross-check of all objects in each K2 campaign field with available variable and binary star catalogs.  164 of 7757 targets selected for observation in campaign 0 and 49 of 21647 targets in campaign 1 were pre-identified as EBs.

In the engineering data-set there are a total of \numTargs\ long-cadence objects observed in addition to \numEngAps\ engineering apertures.  Data were observed in a cadence of 30 minutes and spanning a total of 12 days.

\subsection{Light Curve Extraction from Pixel Data}

For the engineering run of K2, only calibrated pixel data were made available, in contrast to the data-sets released for the original mission which also included extracted light curves. For this work, we have extracted light curves from the pixel data ourselves, using the tools used and presented in, e.g., \citet{Papics13}. We have removed the background flux in the pixels using a low-order spline fit to all available pixels around the targets. The light curves were then constructed by adding up all flux in the pixels around the central pixel that have more than 100 counts. We find this to be a close-to-optimal choice, given that including pixels with less flux will increase the noise and limiting the pixel selection to pixels with higher count levels increases systematic trends.

The extracted lightcurves are detrended to remove any trends, instrumental or astrophysical, not related to the EB signal.  This is done using an iterative sigma-clipping technique to divide by a polynomial fitted to the baseline of the data (see \citealt{Prsa11} for details).

\section{Results: EBs in the K2 Engineering Dataset\label{sec:results}}
\subsection{Manual EB Identification\label{subsec:manual}}

In the K2 engineering target list, 9 objects (60017809, 60017810, 60017812, 60017814, 60017815, 60017816, 60017818, 60017821, 60017822) were identified as previously known EBs.  
One of these (60017818) did not show a clear EB in the 12 days of data, so was excluded, but the remaining 8 were all recovered independently.    

Through manual inspection of all \numTargs\ long-cadence lightcurves, we identified a total of \numManual\ EBs in the K2 engineering dataset (Table~\ref{table:ephems}).  In the original mission we identified EBs through a variety of methods \citep{Prsa11}, but since there were no Threshold Crossing Events (TCEs) released for the engineering dataset, manually inspecting each lightcure was a necessary step in order to test the feasibility of automated detection of eclipsing binary signals in K2 data.  EBs were identified if they showed clear periodic ellipsoidal variation or eclipses in the lightcurves that repeated at least 3 times in the 12 day baseline of the data.  If a lightcurve showed one or two single eclipse events, the EB is included in the list, but ephemerides could not be determined.  Planet Hunters\footnote{\url{http://www.planethunters.org}} \citep{fischer2012} had independently detected and identified several of these EBs as well.

Of these \numManual\, there were \numBlendSets\ sets of nearby targets that exhibited the same period and shape in their lightcurves.  It is likely that we are seeing the same EB signal from a single source bleeding into both apertures.  Due to the large sizes of the apertures in the engineering data, it is difficult to determine the true source of any EB signal, and there is no direct mapping from Kepler ID to stellar objects.  In these cases, the target with the larger amplitude signal was considered the true source and the other target was marked as a blend (false-positive) and removed from the catalog.  A list of these are given in Table~\ref{table:blends}, leaving \numEBs\ manually detected EBs.

The K2 engineering target list, unlike the KIC (Kepler Input Catalog) used for the original \kepler mission and the EPIC (Ecliptic Plane Catalog) used for the K2 campaigns, does not include target object names.  All identified EBs were cross-matched against known sources by their target coordinates with a radius of 1 arcminute.  These nearby sources and their previous characterizations are listed in Table \ref{table:crossmatch}.  We have thus identified \numEBsnew\ previously unknown EBs.

Kepler ID 60017806 was also initially identified as a candidate EB, but is actually a known extrasolar planet (WASP-28b) and was removed from the catalog.

\subsection{EB Ephemerides and Morphologies\label{subsec:ephems}}

\begin{figure}
\centering
\includegraphics{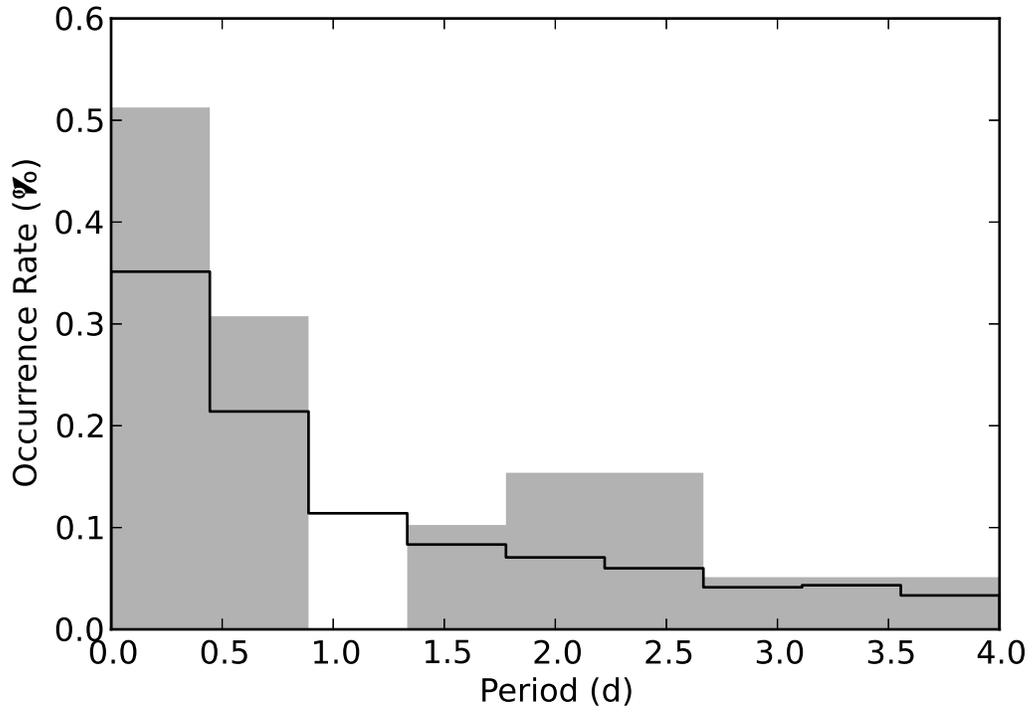}
\caption{Occurrence rate as a function of period for the K2 engineering EBs (gray bars) and EBs from the original \kepler mission (black outline).}
\label{fig:periodhist}
\end{figure}

Ephemerides for the EB systems that exhibited at least 3 subsequent eclipse events are determined by computing a periodogram for each detrended lightcurve using BLS \citep{kovacs2002}, manually adjusting the correct period if necessary, and setting ${\rm BJD}_{0}$ so that the deeper eclipse is placed at zero phase. The ephemerides for all \numEBs\ EBs are listed in Table \ref{table:ephems} and are available online at \texttt{http://keplerEBs.villanova.edu/k2}.  Despite such a small sample size, the distribution in EB orbital periods is consistent with that found from the original mission (Fig.~\ref{fig:periodhist}), with a total detected EB occurrence rate of \fracEBs.

\begin{figure}
\plotone{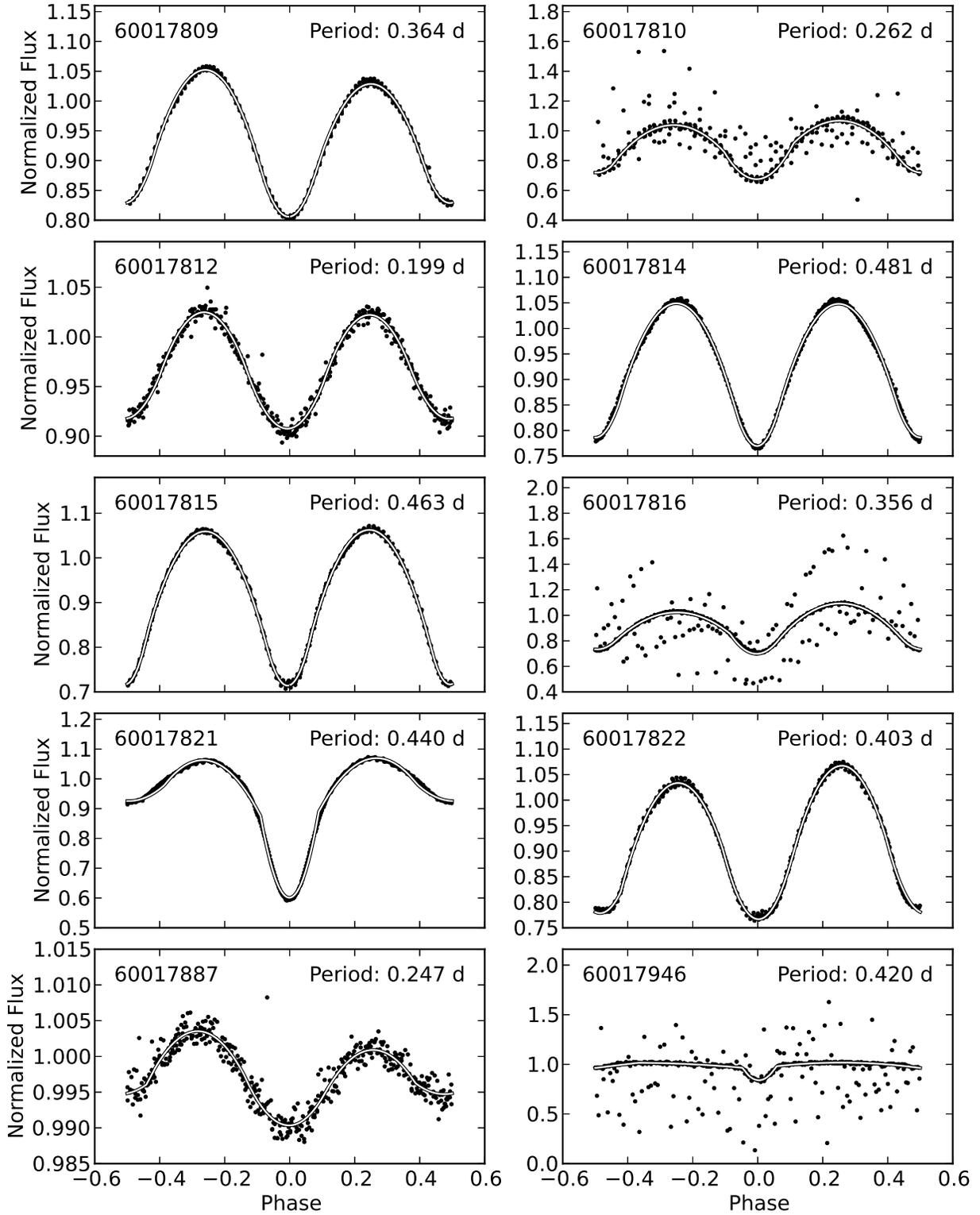}
\caption{Phased, detrended data with polynomial chain fits overplotted, sorted by the orbital period.}
\label{fig:lcs}
\end{figure}

The lightcurves are phase-folded (Fig.~\ref{fig:lcs}) and fit by a chain of four quadratic functions that describe the shape of the phased lightcurve \citep{Prsa11}. This analytic function is then used to determine the morphology, a value between 0 (detached) and 1 (overcontact), using Locally Linear Embedding \citep{Matijevic12}. These values are listed in Table \ref{table:ephems} under the \texttt{morph} column.

\subsection{Test of Automated EB Identification\label{subsec:automated}}

The EBs identified in the K2 dataset provide an initial
benchmark set for newly developed pipelines intended for
automated discovery of EBs from large datasets such as those
that will be provided by the ongoing K2 mission. We applied
the 
Eclipsing Binary Factory (EBF) pipeline 
\citep{Paegert2014,Parvizi}
to the K2 light curves to test its ability to correctly
recover these EBs. 
The EBF correctly recovered 92\% of the manually identified K2 
EBs with at least 90\% confidence in the classification. 
This recovery rate is similar to that obtained 
by the EBF from the original \kepler data set \citep{Parvizi}, suggesting that automated methods such as the EBF are capable of identifying a large sample of EBs in the upcoming K2 campaigns with good
completeness.

\section{Summary and Discussion\label{sec:summary}}

\numEBstext\ eclipsing binaries in the K2 engineering data-set and their ephemerides have been provided. Although the target masks and lightcurve extraction process are different than they were in the original \kepler mission, the developed tools are still applicable and the acquired data are still of high quality for most eclipsing binary science, including all future campaigns of the K2 mission. 

The fraction of EBs identified in the K2 engineering data-set is \fracEBs\, in agreement with the fraction of EBs having periods shorter than 5 days in previous Kepler EB studies \citep{Prsa11,Slawson11,KirkPrep}.

The results of this pilot study show that K2 light curves is a trove of data for identification, classification, and detailed study of EBs along the ecliptic, which include a number of interesting stellar populations (e.g., large numbers of benchmark clusters of various ages) that were not included in the original Kepler footprint \citep{prsa2014}. Visual identification remains an effective approach to identifying EBs with high completeness. However, approaches such as the EBF pipeline \citep{Paegert2014} show good promise for fully automating this search and achieving an equivalent level of completeness.

\acknowledgments
The authors gratefully acknowledge everybody who has made \emph{Kepler}, and especially the K2 mission, possible. 
KEC and KGS gratefully acknowledge support from 
NASA ADAP grant NNX12AE22G.
AP gratefully acknowledges support from the NASA \kepler PSP grant NNX12AD20G. SB is supported by the Foundation for Fundamental Research on Matter (FOM), which is part of the Netherlands Organisation for Scientific Research (NWO).
This research has made use of the SIMBAD database, operated at CDS, Strasbourg, France.

\begin{figure}
\plotone{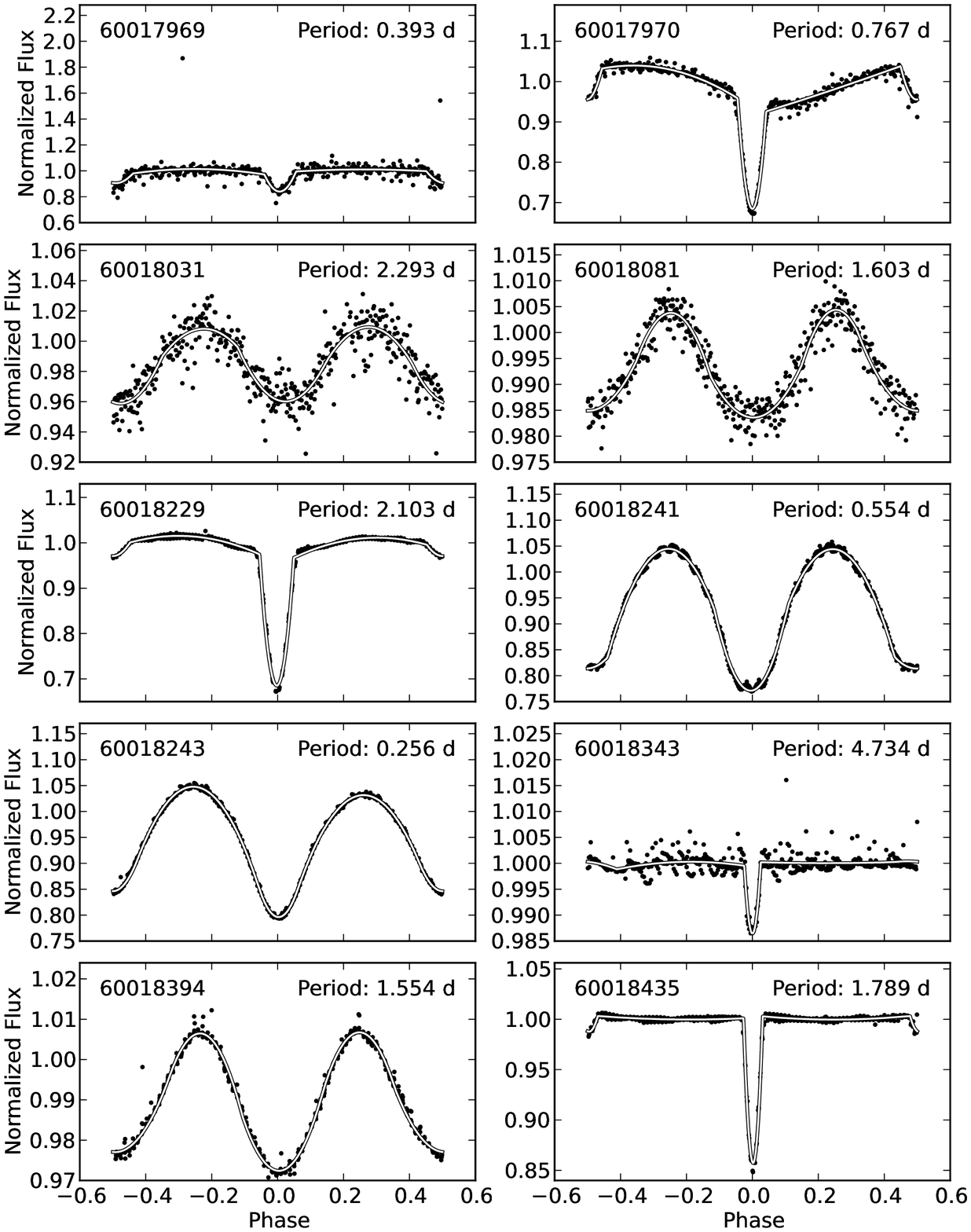}
\caption{Fig.~\ref{fig:lcs} continued.}
\end{figure}

\begin{figure}
\plotone{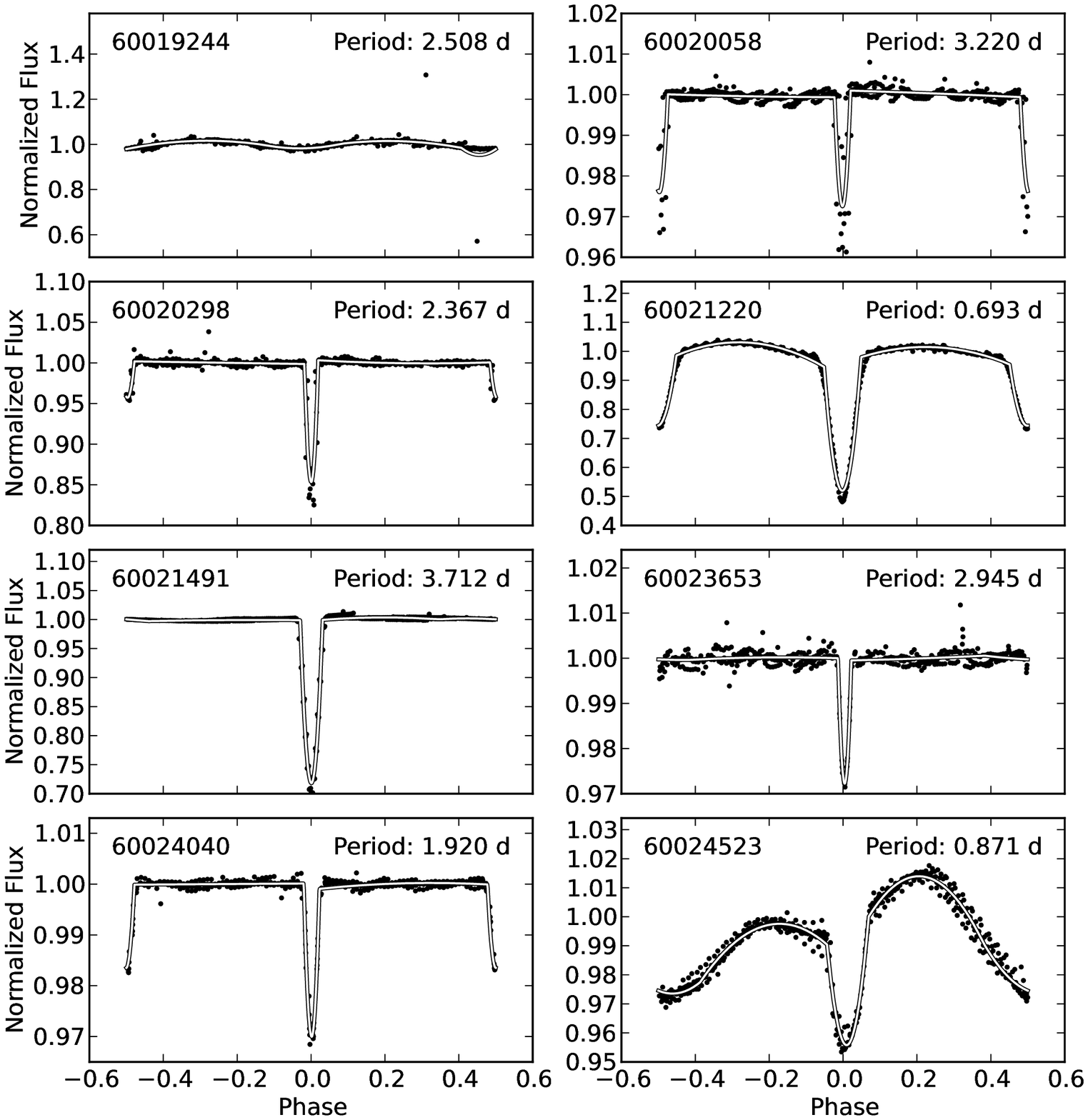}
\caption{Fig.~\ref{fig:lcs} continued.}
\end{figure}

\begin{deluxetable}{rrrrrrr}
\tablecolumns{7}
\tablewidth{0pt}
\tablecaption{Eclipsing Binaries in K2  \label{table:ephems}}
\tablehead{
	\colhead{Kepler ID} & \colhead{Kepler mag} & \colhead{RA (deg)} & \colhead{DEC (deg)} & \colhead{morph} & \colhead{Period (d)} & \colhead{${\rm BJD}_{0} - 2400000$} 
}
\startdata
60017809 & 11.51 & 352.388100 & -3.768842 & 0.87 & 0.363690 & 55001.137420 \\ 
60017810 & 14.53 & 1.157580 & 3.550330 & 0.79 & 0.261511 & 55000.167272 \\ 
60017812 & 16.39 & 4.170960 & -0.156970 & 0.94 & 0.198577 & 54999.926036 \\ 
60017814 & 10.40 & 356.826450 & -8.086691 & 0.80 & 0.481459 & 54998.495542 \\ 
60017815 & 12.00 & 355.528771 & -3.099600 & 0.77 & 0.463411 & 55000.476920 \\ 
60017816 & 13.00 & 352.914004 & -2.701678 & 0.79 & 0.355757 & 55000.900452 \\ 
60017821 & 13.00 & 355.093408 & -7.796992 & 0.65 & 0.439926 & 54999.687165 \\ 
60017822 & 11.30 & 352.818348 & -5.371712 & 0.90 & 0.403352 & 54999.062728 \\ 
60017887 & 10.51 & 352.531675 & 1.434328 & 0.90 & 0.247156 & 54999.642249 \\ 
60017946 & 17.25 & 357.040958 & -0.532353 & 0.56 & 0.420497 & 54998.017630 \\ 
60017969 & 19.15 & 356.424312 & 0.439217 & 0.54 & 0.393173 & 55001.429657 \\ 
60017970 & 15.74 & 351.671617 & 0.795622 & 0.54 & 0.766959 & 54999.282561 \\ 
60018031 & 18.61 & 0.786213 & 0.130258 & 0.90 & 2.293002 & 55007.101680 \\ 
60018081 & 13.06 & 353.644290 & -1.326940 & 0.97 & 1.602950 & 54999.175332 \\ 
60018229 & 12.57 & 1.362500 & 4.806667 & 0.55 & 2.103210 & 55000.633484 \\ 
60018241 & 12.54 & 356.162500 & -1.810000 & 0.82 & 0.553764 & 54999.867934 \\ 
60018243 & 13.33 & 359.750000 & -9.526667 & 0.77 & 0.256241 & 55000.279596 \\ 
60018343 & 10.05 & 2.241575 & 2.945010 & 0.36 & 4.734390 & 54999.962785 \\ 
60018394 & 10.22 & 354.033199 & -6.232208 & 0.94 & 1.553901 & 55006.906322 \\ 
60018435 & 10.37 & 5.163989 & -5.143139 & 0.40 & 1.788873 & 55000.404974 \\ 
60019244 & 14.40 & 359.491080 & -3.689460 & 0.67 & 2.507734 & 54988.973929 \\ 
60019950 & 14.80 & 354.970700 & 1.983330 & \nodata & \nodata & 55000.063356 \\ 
60020058 & 14.80 & 356.159940 & -8.852300 & 0.38 & 3.220334 & 55001.232252 \\ 
60020298 & 14.90 & 354.698130 & -7.806650 & 0.29 & 2.366542 & 54999.069880 \\ 
60021220 & 15.47 & 356.476390 & -0.525299 & 0.55 & 0.692561 & 55000.451393 \\ 
60021491 & 11.41 & 0.446743 & -3.168466 & 0.41 & 3.711619 & 55001.674674 \\ 
60021545 & 10.60 & 0.806291 & -3.911404 & \nodata & \nodata & 55000.315793 \\ 
60023653 & 10.33 & 355.034714 & -2.480564 & 0.27 & 2.944828 & 55000.065586 \\ 
60024040 & 10.62 & 357.762031 & -2.594677 & 0.33 & 1.919605 & 55000.113169 \\ 
60024244 & 12.22 & 358.906979 & -4.369421 & \nodata & \nodata & 55003.999466 \\ 
60024523 & 11.04 & 3.678608 & -5.215159 & 0.77 & 0.871013 & 54999.948571 \\ 
\enddata
\end{deluxetable}

\begin{deluxetable}{rl}
\tablecolumns{2}
\tablewidth{0pt}
\tablecaption{Blended EBs in K2. \label{table:blends}}
\tablehead{
    \colhead{Kepler ID (EB)} & \colhead{Kepler ID (blend)}
}
\startdata
60017809 & 60023285 \\
60017815 & 60018240 \\
60017816 & 60042608 \\
60017822 & 60023349 \\
60018081 & 60017828 \\
60024523 & 60024522 \\
\enddata
\end{deluxetable}

\begin{deluxetable}{rl}
\tablecolumns{2}
\tablewidth{0pt}
\tablecaption{Cross-matched identifications for EBs in K2. \label{table:crossmatch}}
\tablehead{
    \colhead{Kepler ID} & \colhead{Objects within 1 arcmin and their Simbad classifications}
}
\startdata
  60017809\tablenotemark{1} & 2MASS J23293314-0346078 (Candidate EB*); 1RXS J232933.9-034601 (X); \\
  60017810\tablenotemark{1} & 1SWASP J000437.82+033301.2 (Candidate EB*); \\
  60017812\tablenotemark{1} & 2MASS J00164102-0009251 (low-mass*); \\
  60017814\tablenotemark{1} & V* EL Aqr (EB*WUMa); \\
  60017815\tablenotemark{1} & TYC 5255-370-1 (Candidate EB*); \\
  60017816\tablenotemark{1} & 2MASS J23313936-0242060 (Candidate EB*); \\
  60017821\tablenotemark{1} & NSVS  11904371 (Candidate EB*); \\
  60017822\tablenotemark{1} & TYC 5257-616-1 (Candidate EB*); 1RXS J233116.9-052239 (X); \\
  60017887 & 2MASS J23300759+0126037 (pMS*); \\
  60017946 & SDSS J234809.83-003156.4 (low-mass*); \\
  60017969 & SDSS J234541.83+002621.1 (low-mass*); \\
  60017970 & SDSS J232641.19+004744.1 (low-mass*); \\
  60018031 & SDSS J000308.69+000749.0 (low-mass*); \\
  60018081 & V* EQ Psc (V*); \\
  60018229 & TYC    4-517-1 (Star); \\
  60018241 & NSVS  11906468 (Candidate EB*); \\
  60018243 & \nodata \\
  60018343\tablenotemark{2} & TYC    4-331-1 (Star); \\
  60018394 & BD-07  6054 (Star); \\
  60018435 & BD-05    43 (Star); \\
  60019244 & \nodata \\
  60019950 & \nodata \\
  60020058 & 2MASS J23443838-0851082 (Star); HD 222891 (Candidate EB*); 1RXS J234438.7-085054 (X); \\
  60020298 & PB  7745 (Star); \\
  60021220 & \nodata \\
  60021491 & TYC 4666-383-1 (Star); \\
  60021545 & TYC 4666-518-1 (Star); \\
  60023653 & BD-03  5686 (Star); \\
  60024040 & TYC 5256-76-1 (Star); \\
  60024244 & TYC 5256-1076-1 (Star); \\
  60024523 & \nodata \\
\enddata
\tablenotetext{1}{Kepler ID is listed as an EB in K2 Engineering Target List}
\tablenotetext{2}{Identified by \citet{Poleski10} (Table 1, line 5) as an SB1 EB with a period of 4.72277 d}
\end{deluxetable}

\end{document}